\documentclass[review]{elsarticle}
\usepackage{lineno,hyperref}
\usepackage{graphicx}
\usepackage{amsmath}
\usepackage{amssymb}
\usepackage{pifont}
\usepackage{booktabs,tabularx}
\usepackage{caption,bm}
\usepackage{subcaption}
\usepackage[normalem]{ulem}
\usepackage{color,soul}
\modulolinenumbers[5]
\makeatletter \renewcommand{\@dotsep}{10000} \makeatother 

\biboptions{authoryear}
\newcommand{\fref}[1]{Fig.~\ref{#1}}
\newcommand{\tref}[1]{Table.~\ref{#1}}
\newcommand{\req}[1]{Eq.~(\ref{#1})}
\begin{document}

\begin{frontmatter}

\title{On growth, buckling, and rupture of aneurysms}

%% Group authors per affiliation:
\author[mymainaddress]{Masoud Hejazi}
\author[mymainaddress]{A. Srikantha Phani\corref{mycorrespondingauthor}}
\cortext[mycorrespondingauthor]{Corresponding author \newline phone: +1 (604) 822-6998 \newline fax: +1 (604) 822-2403 \newline email: srikanth@mech.ubc.ca}

\address[mymainaddress]{Department of Mechanical Engineering, 6250 Applied Science Lane, University of British Columbia, Vancouver, B.C, Canada V6T 1Z4}

\begin{abstract}
%compare and contrast two models. failure criteria (gent and ogden for hyperleastic), cauchy for GOH. stretch inversion, exponential stiffening, propagation pressure is slighlty lower. Sequence of bulging and buckling.

Aneurysms are localized bulges of arteries, and they can rupture with fatal consequences. Complex mechanobiological factors preclude \emph{in vivo} testing to assess the rupture risk of an aneurysm, and size based criteria are often used in clinical practice to guide surgical interventions. It is often found that tortuous and buckled aneurysms can exceed in size recommended for surgical intervention, and yet do not rupture. This study addresses why buckled aneurysms exhibit this intriguing behaviour by combining \emph{in vitro} inflation experiments on hyperelastic tubes with finite element calculations. Using a biologically relevant material model for an arterial wall, we show that buckled aneurysms can grow in size without rupturing under favourable arterial pre-tensions. Stretch reversal phenomenon exhibited by arteries governs whether buckling or bulging occurs first. Exponential stiffening favours the axial propagation of an aneurysm instead of radial growth in size. The choice of failure criteria based on Ogden's strain energy function, Gent's first stretch invariant, and Cauchy stress are discussed. Failure maps incorporating post bifurcation (bulging and buckling) response are constructed to delineate the regimes of growth, buckling and rupture of an aneurysm.

\end{abstract}
\begin{keyword}
Aneurysm, Buckling, Rupture, Finite Element Analysis~\\
%Word Count: 3436
\end{keyword}
\end{frontmatter}

%\linenumbers

%%%%%%%%%%%%%%%%%%%%%%%%%%%%%%%%%%%%%%%%%%%%%%%
%%%%%%%%%%%%%%%%%%%%%%%%%%%%%%%%%%%%%%%%%%%%%%%
\section{Introduction}
An aneurysm is a balloon-like bulge within the wall of a blood vessel,  commonly observed in the abdominal and the thoracic regions of aorta, due to local  degradation in mechanical properties of an artery~\citep{sakalihasan2018}. Biomechanics of aneurysms has been reviewed in~\citep{vorp2007,sakalihasan2018}, arterial wall modelling in~\citep{holzapfel2010rev,humphrey2013,fung2013}. An increasing incidence rate of aneurysms, ranging from 5.6 to 10.6 per 100,000 persons/year, has been observed in the past decade  in developed countries~\citep{acosta2006}. Mortality rate for a ruptured aortic aneurysm {can be} up to 90\%, especially when it occurs outside of a hospital~\citep{sampson2014}. Hence, estimating the rupture risk for AAAs is  critical after an aneurysm is detected.   Clinical guidelines rely on the maximum diameter of an aneurysm to estimate the rupture risk~\citep{muluk2017,lazaris2019}. However, some AAAs are found unruptured despite their maximum size exceeding the minimum recommended size (5.5 cm) for surgical interventions~\citep{lazaris2019,fillinger2004}. Tortuous and buckled AAAs  whose size far exceed the guidelines have been observed without rupture~\citep{fillinger2004,hejazi2021,boyd2016,boyd2021,sall2021}. Arterial stability with aneurysm~\citep{lee2014mec} and buckling without aneurysms~\citep{liu2019} has also been studied, among others. However, the rupture of buckled aneurysms remains to be addressed.

Recently, the hypothesis that buckling offers a fail safe mode for AAAs has been tested through controlled \emph{in vitro} experiments on hyperelastic tubes and numerical modelling~\citep{hejazi2021} by treating the walls as isotropic and hyperelastic membranes. Pre-tension has been identified as the key factor that governs whether buckling occurs or not. It has been reported that, under low pre-tensions (arterial pre-stress) an aneurysm buckles and grows in size without rupture. Buckling of an aneurysm is shown to arise from the compressive stresses developed outside the bulged region which overcome the arterial pre-tension. Hyperelastic models may be justified on the basis of the degradation of elastic fibres due to  aneurysm's pathogenesis at incipient rupture~\citep{niestraska2019}. Bulge formation and other bifurcation phenomena in hyperelastic cylindrical tubes have been  studied extensively using approximate membrane theories for thin walled tubes~\citep{haughton1979a,haughton1979b,fu2012}, diffuse interface models~\citep{lest2018,lest2020}, and for thick walled cylindrical tubes~\citep{fu2016,wang2018,wang2019,ye2020}.  Hyperelastic models have also been used in finite element calculations~\citep{lanne1992,gonccalves2008} to study bulge formation in cylindrical tubes. While a phenomenological and qualitative understanding  of aneurysm formation and growth can be obtained using hyperelastic models, they may not be suitable to describe the complex material response of a fiber-reinforced, multi-layered, and anisotropic arterial wall~\citep{holzapfel2019}.  Attempts have been made to use more complex material models in finite element calculations~\citep{alhayani2013,demirkoparan2017,font2021,topol2021,dehghani2019}. These studies indicate the asymmetry of aneurysm shape~\citep{topol2021,font2021} due to residual stresses, buckling without bulging~\citep{dehghani2019}, competition between radial expansion and axial propagation of a bulge~\citep{alhayani2014}, and the effect of fiber reinforcement on the initiation of the aneurysm~\citep{demirkoparan2017}. However, these computational studies do not address the rupture in the post bulging regime;  the role of buckling in preventing it; the influence of arterial pre-tension on the occurrence of buckling, bulging and rupture. Significantly, features such as stretch reversal and exponential stiffening are absent in hyperelastic material models and their influence on aneurysm evolution needs to be clarified.

This study aims to understand the influence of biologically relevant constitutive material laws on the biomechanics of growth and rupture of the abdominal aortic aneurysm (AAA).  To do so, we use the Gasser-Ogden-Holzapfel (GOH) constitutive model of arteries~\citep{gasser2006} to account for exponential stiffening and stretch reversal~\citep{sommer2010}.  Experimental protocols and finite element analyses are described in Section 2. Section 3 discusses results and factors governing the growth and rupture of an aneurysm, culminating in  failure maps.  Clinical implications and limitations are discussed. Cncluding remarks and avenues for further investigation  are provided in Section 4.

\section{Methodology}
We first present the experimental methodology, followed by finite element calculations. Experiments on hyperelastic tube models will be used to validate the finite element methodology, which will then be extended to include biologically relevant material models. 
\subsection{Experiments on hyperelastic  cylindrical tubes}
\begin{figure}[!h]
	\centering
	\includegraphics[width=\textwidth]{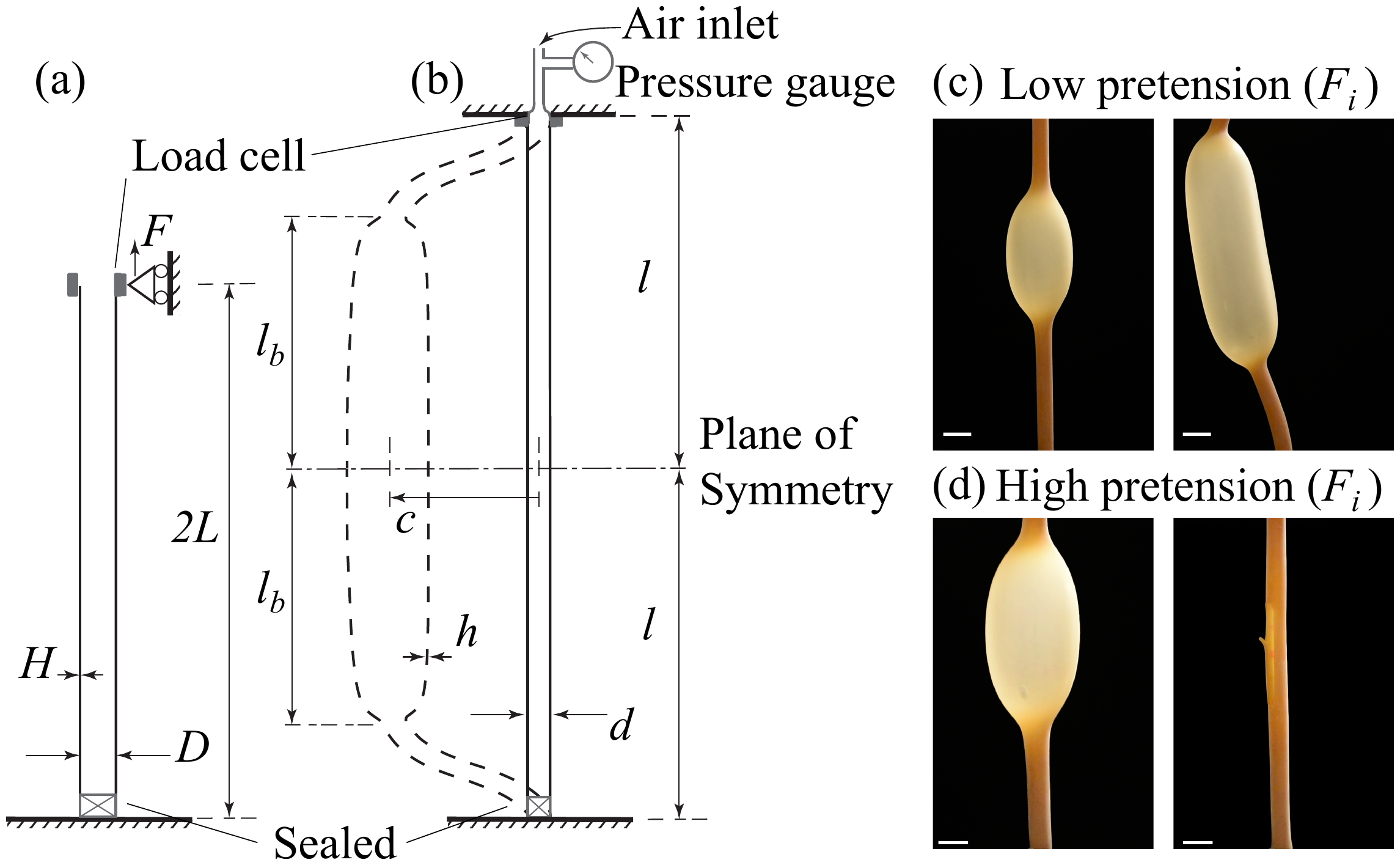}
	\caption{Experimental setup for inflation test of pre-streched rubber tube with fixed-fixed boundary conditions, $L$ (initial half-length), $D$ (initial diameter), $H$ (initial thickness), $F$ (force applied/measured by Intron), $l$ (stretched half-length), $l_b$ (bulge length), $d$ (deformed diameter), $c$ (buckling amplitude), $F_{i}$ (longitudinal tension), and $h$ (deformed thickness). (a) Longitudinal tension is applied at the upper end to reach the required pre-stretch ($\lambda_{i}$). (b) Deformed (dashed) and undeformed configurations during inflation subjected to fixed-fixed ends. (c) Bulge initiation and buckling at low longitudinal tension (pre-stretch). (d) Bulge initiation and rupture at high longitudinal tension. The scale bar in white is 2~cm.}
	\label{Fig_1}
\end{figure}

Inflation tests on finite latex rubber (hyperelastic) tubes are conducted as illustrated in~\fref{Fig_1}. The purpose of these tests is to understand the deformation phenomena of buckling and kinking, and their influence  on the axial and radial growth of a bulge. Here, the emphasis is on post bulge regime, and the influence of material models on this regime that has  not been investigated in the literature.  Material and geometric properties of the tubes used are given in~\tref{Table_1}.  A nominally straight cylindrical rubber tube is fixed at its bottom, and the top end is attached to an Instron  (model 5969) tensile tester's jaw.  A known pre-tension (pre-stretch) is applied by moving the jaw  upward and then holding it fixed at a  prescribed displacement throughout the inflation test. The load cell reads the axial force ($F$) in this displacement control test. The air inlet is attached to the top end of the tube and a pressure gauge reads the injection pressure. A DSLR camera (Nikon D7000) records the deformation of the tube during air  injection and a pressure transducer records the internal pressure. An edge detection algorithm in  MATLAB{$^ \copyright$} image processing toolbox is used to obtain the bulged profile for further analysis. A number of experiments (39 in total) are performed for different initial pre-tensions, $F_i$, and  aspect ratios by changing the initial lengths of the tube. The initial diameter and the wall thickness are held fixed in all the experiments.  Depending on the initial pre-tension ($F_i$) two possible outcomes of the inflation tests are shown in~\fref{Fig_1}(c) and(d), to be   discussed later in Section 3.1.

\subsection{Finite element analyses}

%Recently, finite element analysis (FEA) has become more popular to study the bulge formation in hyperelastic tubes~\citep{alhayani2014,dehghani2019,font2021}. Compared to asymptotic analytical solutions, FEA gives the freedom to explore the post bifurcation behaviour with higher accuracy, minor assumptions on the solution and less computation time. Furthermore, it is easier to implement the effect of material properties, and pre-stretch, since the governing equations remain the same and only numerical calculation requires to be carried on for a new set of material, geometry, and loading conditions.
Finite element analysis (FEA) offers the ability to systematically vary the material and geometric parameters of the inflation test. We conduct  FEA in two parts. In the first part, virtual inflation tests on  cylindrical hyperelastic tubes are performed. Initial tension is varied in these simulations. The purpose of these simulations is to  verify the experimental observations of buckling in the post bulge regime as well as  validate the modelling procedure. In the second part, we use this FEA procedure with a more biologically relevant GOH material model for the wall.

The FEA has been conducted in three stages. In experiments, all the displacement degrees of freedom have been constrained at the lower end, and at the top end of the tube, as shown in~\fref{Fig_1}(b). We can exploit  symmetry of the boundary conditions by analyzing only one half of the tube. Note that this symmetry still allows the aneurysm to to grow radially asymmetric. The top end of the tube with half length, $l$, as shown in~\fref{Fig_1}(b), coincides with the crown of the bulge and it is only constrained for the axial displacement to reflect symmetry.  The plane of symmetry is also indicated. We use ABAQUS$^\copyright$ Linear Perturbation Analysis (LPA) and Riks arc length method  solvers to obtain the critical bifurcation modes and post bifurcation solution, respectively~\citep{aba2012,crisfield1983}. In stage 1 of the FEA, LPA has been performed to obtain the unstable mode shapes of a cylindrical tube subjected to  a  normalized unit value of  prescribed axial stretch and internal pressure. The global nodal imperfection has been introduced based on the first two circumferential modes (global bulge shape) and the first axial mode. It should be noted that due to manufacturing defects, these imperfections naturally exist in the tubes subjected to the inflation test in Section 2.1. In stage 2, the desired pre-stretch is applied and held constant for various prescribed values, identical to that in experiments on rubber tubes. Finally, the inflation test is simulated  by  increasing the internal pressure. Both stages 2 and 3 use Riks method.

%, and the Riks' method to capture the post bifurcation regime.  

Linear shell elements with reduced integration (S4R) were used~\citep{aba2012}. We used 31 elements for the circumference and 199 elements along the length of the tube (6169 total). Element size and amplitude of imperfection are two important simulation parameters for the bifurcation analysis. For the Riks method, global nodal imperfection of order of $10^{-3}$, normalized with respect to element size, has been introduced for the first two circumferential modes (global bulge shape) and the first axial mode (buckling mode). The number of elements and the  normalized (with respect element size) modal amplitude of imperfection  ($10^{-3}$) have been chosen based on the convergence of the critical pressure associated with bulging instability. 

%The element size is chosen based on the convergence of critical pressure associated with bulge formation. The normalized imperfection amplitude ($10^{-3}$) is arrived at by following the same convergence criterion ofr the critical pressure. 

\subsection{Material models}
Two material models are considered in the finite element analyses: isotropic hyperelastic wall material and anisotropic  GOH model. For the hyperelastic model, we use two failure criteria, respectively based on Ogden's strain energy function (SEF) and Gent's criterion based on the first stretch invariant. For the GOH model we Cauchy stress based failure criterion following~\citep{sherifova2019biomechanics}.

The Ogden' SEF ($W_{\text{O}}$) for a cylindrical coordinate with plane stress can be expressed as  
\begin{equation} \label{ogden}
\begin{array}{c}
W_{\text{O}}=\displaystyle\sum_{i=1}^{3}\frac{s_i}{\alpha_{i}}(\lambda_{1}^{\alpha_i}+\lambda_{2}^{\alpha_i}+\lambda_{3}^{\alpha_i}-3), \\
\lambda_{1} = \lambda_{z}, ~\lambda_{2} = \lambda_{\theta}, ~\lambda_{3}=(\lambda_{z}\lambda_{\theta})^{-1}, 
\end{array}
\end{equation}
where ($\lambda_z$ and $\lambda_\theta$) are principal stretches in cylindrical coordinate system for an axisymmetric membrane. The material parameters are given in~\tref{Table_1} based on our previous experimental calibration~\citep{hejazi2021}. We used two different rupture criteria: (1) The maximum strain energy density based on Ogden's model in~\req{ogden}; (2) the maximum in-plane first stretch invariant based on Gent's material model~\citep{gent1996}, which reads 
\begin{equation}\label{gent}
	\begin{array}{c}
	I = \lambda_{1}^{2} + \lambda_{2}^{2} + \lambda_{3}^{2} -3, \\
	\lambda_{1} = \lambda_{z}, ~\lambda_{2} = \lambda_{\theta}, ~\lambda_{3}=(\lambda_{z}\lambda_{\theta})^{-1}.
	\end{array}
\end{equation}
First, we use the calibrated material parameters for Ogden’s strain energy function (SEF) in~\req{ogden} to validate our FEA with experiments. Next, with  identical assumptions as above, namely, axisymmetric membrane and plane stress, the Gasser-Ogden-Holzapfel (GOH) model's SEF ($W_{\text{H}}$) can be expressed as
\begin{equation} \label{holz}
	\begin{array}{c}
		W_{\text{H}}=\displaystyle\frac{c_{10}}{2}(I_{1}-3)+\displaystyle\frac{k_1}{k_2}\left(\text{exp}[k_{2}({\kappa}I_{1}+(1-3\kappa)I_{4}-1)^2]-1\right),\\
		I_{1}=\lambda_{z}^{2}+\lambda_{\theta}^{2}+(\lambda_{z}\lambda_{\theta})^{-2},~I_{4}=\lambda_{z}^{2}\sin^2{\gamma}+\lambda_{\theta}\cos^2{\gamma},
	\end{array}
\end{equation}
where, $c_{10}$, $k_1$, and $k_2$ are the material parameters regarding the contribution of strain energy in the matrix and fibers. The structural parameters are $\kappa$ and $\gamma$ represent the in-plane fiber gaussian distribution and fiber orientation with respect to circumferential direction ($\theta$), respectively~\citep{gasser2006}. The first term in the above equation containing $c_{10}$ can represent the strain energy stored in the extra cellular matrix and the second term is the strain energy stored in reinforcing fibers (collagen). $k_1$ and $k_2$ respectively correspond to the volume fraction and stiffness of collagen fibers.  The parameters used in the computations are given in~\tref{Table_1} baed on \emph{ex-vivo} calibration test on AAA tissue samples~\citep{niestraska2019}.
\begin{table}[!h]
	\caption{Geometric and material properties of the tubes in experiment and FEA based on \req{ogden} and \req{holz}}\label{Table_1}
	\vspace{-3mm}
	\begin{tabular}{p{4.6cm}p{6.6cm}}
		\hline
		Internal diameter, $D$& 6.5 $mm$\\
		Length, $2L$& 10~-~20 $cm$\\ %10--20 $cm$
		Tube wall thickness, $H$ &  1.5 $mm$\\
		Ultimate uniaxial stretch, $\lambda_{ut}$& $7.85\pm0.8$ \\
		Ogden's SEF parameters& $\alpha_{1}=1.04$; $\alpha_{2}=1.33$; $\alpha_{3}=4.46$ \\
		&   $s_{1}=704.8$; $s_{2}=373.2$; $s_{3}=2.8$ ($s_i$ in kPa) \\
		GOH's SEF parameters& $c_{10}=5.9$~kPa; $k_{1}= 18.2$~kPa; $k_{2}= 17.7$ \\
		&   $\kappa=0.242$; $\gamma=60^{\circ}$ \\
		\hline
	\end{tabular}
\end{table}
%%%%%%%%%%%%%%%%%%%%%%%%%%%%%%%%%%%%%%%%%%%%%%%
%%%%%%%%%%%%%%%%%%%%%%%%%%%%%%%%%%%%%%%%%%%%%%%
\section{Results and discussion}
In this section we first validate the FEA modelling procedure by comparing it with experiments and then use the FEA subsequently for different material models, geometric parameters, and pre-tensions. 

 % ** Delete? We discuss the bulge radial, axial growth, and rupture prediction (based on Gent’s invariant criterion and maximum strain energy density). A failure map based on maximum diameter ($d_\text{max}$) and buckling amplitude ($c$) is sketched. Afterwards, using Holzapfel’s strain energy function, \req{holz}, for fiber reinforced hyperelastic materials; widely used for arterial tissue; we discuss the bulge formation, growth, buckling, and rupture. Finally, the effect of material parameters for Holzapfel’s SEF on the failure map is examined. 

%Following the initial uniform inflation, the critical pressure is reached at the onset of bulge formation. By continuing the inflation, the localized bulge is formed, and the internal pressure drops suddenly producing a bifurcation point on the pressure-volume curve. The propagation of the bulge has two stages, radial propagation and axial propagation. During the axial propagation, the maximum diameter, known as aneurysm diameter ($d_\text{max}$) remains approximately constant, while the length of the bulge ($l_b$) increases leading to either buckling or rupture based on the initial pre-tension (see~\fref{Fig_1}(c) and(d)).
%***
\subsection{Rubber tube: modeling and experiments}

\begin{figure}[!h]
	\centering
	\includegraphics[width=\textwidth]{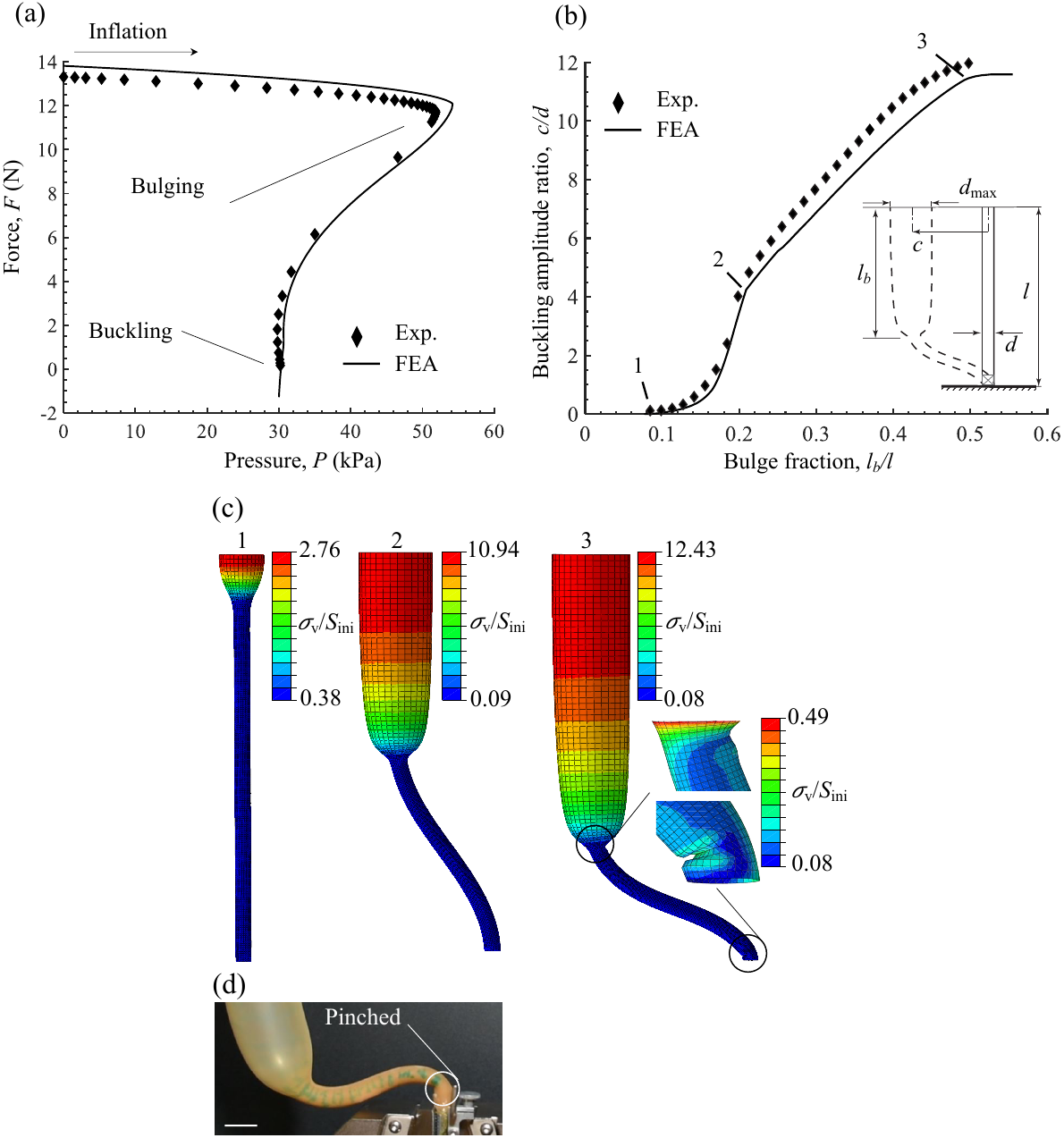}
	\caption{Finite element analysis versus experiments on hyperelastic tubes. (a) Longitudinal force versus internal pressure during inflation. The onset of bulging and buckling are associated with a drop in the internal pressure and a drop in the longitudinal force, respectively. (b) Buckling amplitude growth versus bulge axial growth showing three regimes: (1-2) buckling initiation, (2-3) flexural stiffening due to bulge axial growth and reduction in length of uniform section, (3-end) buckling amplitude limited by local kinking. (c) FEA stress distribution associated with points 1 to 3 in (b) and shear stress at the neck of the the bulge and bottom fixed end subjected to kink. (d) Small kink formation at the bottom fixed end of the tube during experiment.}
	\label{Fig_2}
\end{figure}

With fixed-fixed boundary conditions, a pre-stretched tube subjected to internal pressure undergoes three stages of deformation as shown in~\fref{Fig_2}(a): (1) uniform expansion, (2) radial expansion and axial growth of the aneurysm, (3) buckling and/or rupture. During the uniform expansion, the internal pressure increases monotonically until a global long-wavelength bulge forms with a relatively small amplitude. The formation of the long-wavelength bulge is associated with a reduction of pressure and it rapidly localizes forming the so-called aneurysm, or a bulge. Formation of the localized bulge leads to a sudden pressure relaxation. As the bulge region invades the straight portions of the tube, the axial force drops due to load relaxation. A compressive region forms outside the bulged region, since both ends are constrained, lateral buckling occurs. If the pre-stretch is high enough the tube may rupture without buckling. This is because the decrease in axial force due to an axially propagating bulge is insufficient to overcome the pre-tension.   For the case where buckling follows bulge formation, as shown in~\fref{Fig_2}(a),  the pressure-force curves show a good agreement between the experiments and the FEA. The FEA shows a slightly stiffer response due to slight differences in the material properties and  discretization. In the post buckling regime, shown in~\fref{Fig_2}(b), the normalized buckling amplitude ratio ($c/d$) versus bulge length fraction ($l_{b}/l$) plot shows a similar level of agreement. In the early stages of buckling, the lateral displacement increases rapidly, but with the increasing buckling amplitude ($c$), we observe a hardening behaviour that leads to a plateau (indicated by 3 in~\fref{Fig_2}(b))  in FEA. This deviation arises from the formation of a localized kink at the bottom end and the neck of the bulge (See~\fref{Fig_2}(c)). Compared to FEA, the kink observed in the experiment, see\fref{Fig_2}(d)), is not as strong. Having verified the validity of FEA, we now can vary material model and other parameters.

%Another source of deviation is attributed to the severe mesh distortion and possible shear locking in linear-quadratic shell elements despite the reduced integration scheme followed. Overall, FEA is capable of reproducing the critical bulge initiation and post critical bulge growth and buckling response observed in the experiments. Having verified the validity of FEA, we now can vary material model and other parameters.

\begin{figure}[!h]
	\centering
	\includegraphics[width=\textwidth]{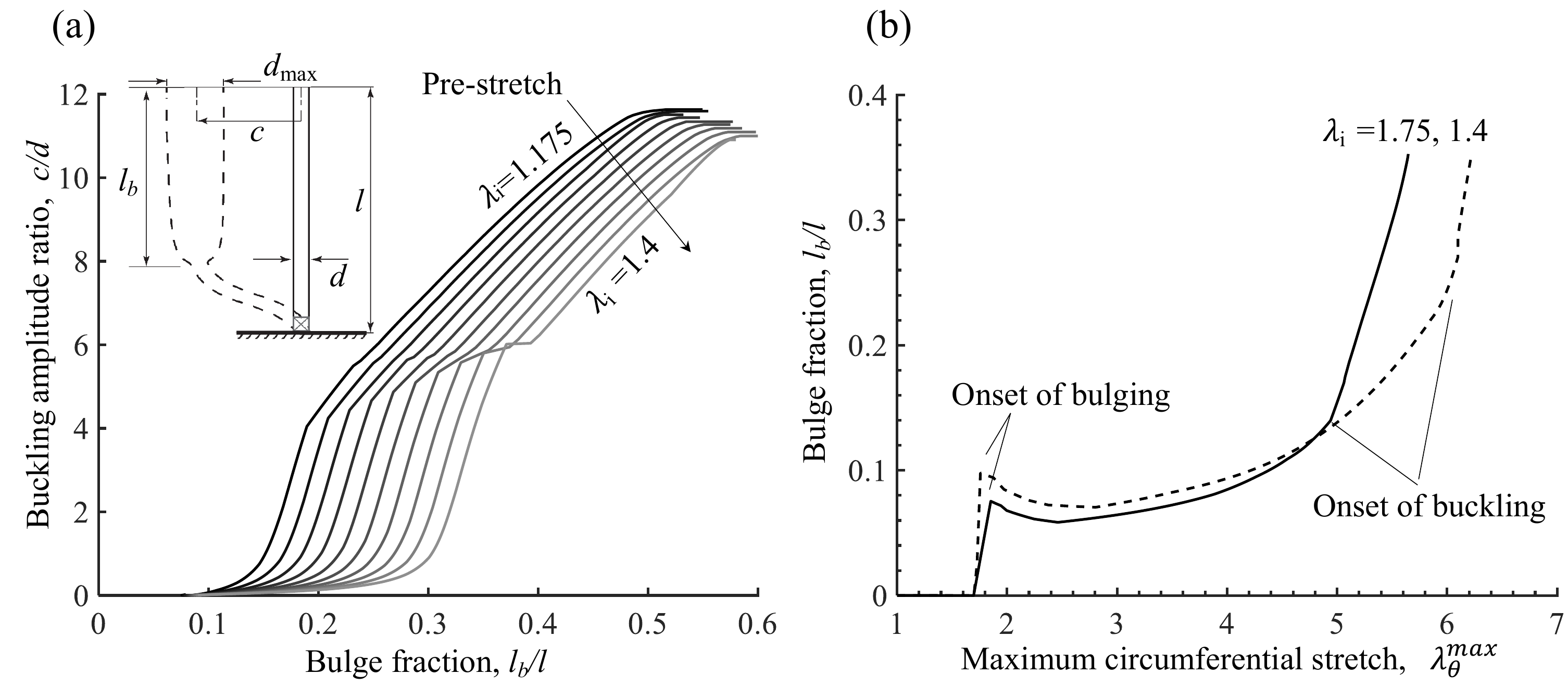}
	\caption{Effect of longitudinal pre-stretch on the growth of $l_b$ (axial length of the bulge) and $c$ (buckling amplitude); $\lambda_{\theta}^{\text{max}}=d_{\text{max}}/d$. (a) By increasing tension, buckling initiation requires larger bulge length and flexural stiffening is more significant due to a shorter length of the uniform section. (b) The long wave bulge profile ($l_{b}/l < 0.1$) is followed by localized bulge formation and rapid axial growth due to buckling. With higher longitudinal pre-stretch the length of the long wave is larger and the buckling initiates at larger diameter/circumferential stretch.}
	\label{Fig_4}
\end{figure}

The effect of pre-stretch on the progression of the buckling and bulge growth is shown in ~\fref{Fig_4}. Here,  the normalized buckling amplitude is plotted against the fraction of the length of the tube, $l_b/l$, occupied by the bulge. With larger axial propagation, the uniform section of the tube becomes shorter, and the end kink  has a more predominant effect on hardening transition in buckling amplitude development (see the transition region in~\fref{Fig_4}(a)). The axial and radial propagation of a bulge is shown in~\fref{Fig_4}(b). At the onset of bulge propagation, the uniform expansion is followed by the formation of a long-wavelength bulge, which then becomes localized by growing in radial size. The radial propagation of the localized bulge gradually turns into axial propagation until the buckling sets in. With buckling, the radial propagation  declines and the bulge length development accelerates (\fref{Fig_4}(b)). This behaviour is a result of relaxation of axial force due to buckling, which allows the bulge to easily propagate along the length of the tube. However, with a higher pre-stretch, the bulge needs to invade longer until the buckling (and hence relaxation) develops. This means that the onset of buckling is delayed in~\fref{Fig_4}(b). 
%At even higher values of pre-stretches (not shown) buckling is prevented and the aneurysm ruptures by radially growing in size.
%Accordingly, the bulge starts its radial growth with a monotonic increase in diameter. The radial growth continues until a steady diameter is reached, after which, the axial growth takes place and the bulge propagates along the length of the tube with a marginal increase of radial size, as shown in~\fref{Fig_2}(b). 

\begin{figure}[!h]
	\centering
	\includegraphics[width=\textwidth]{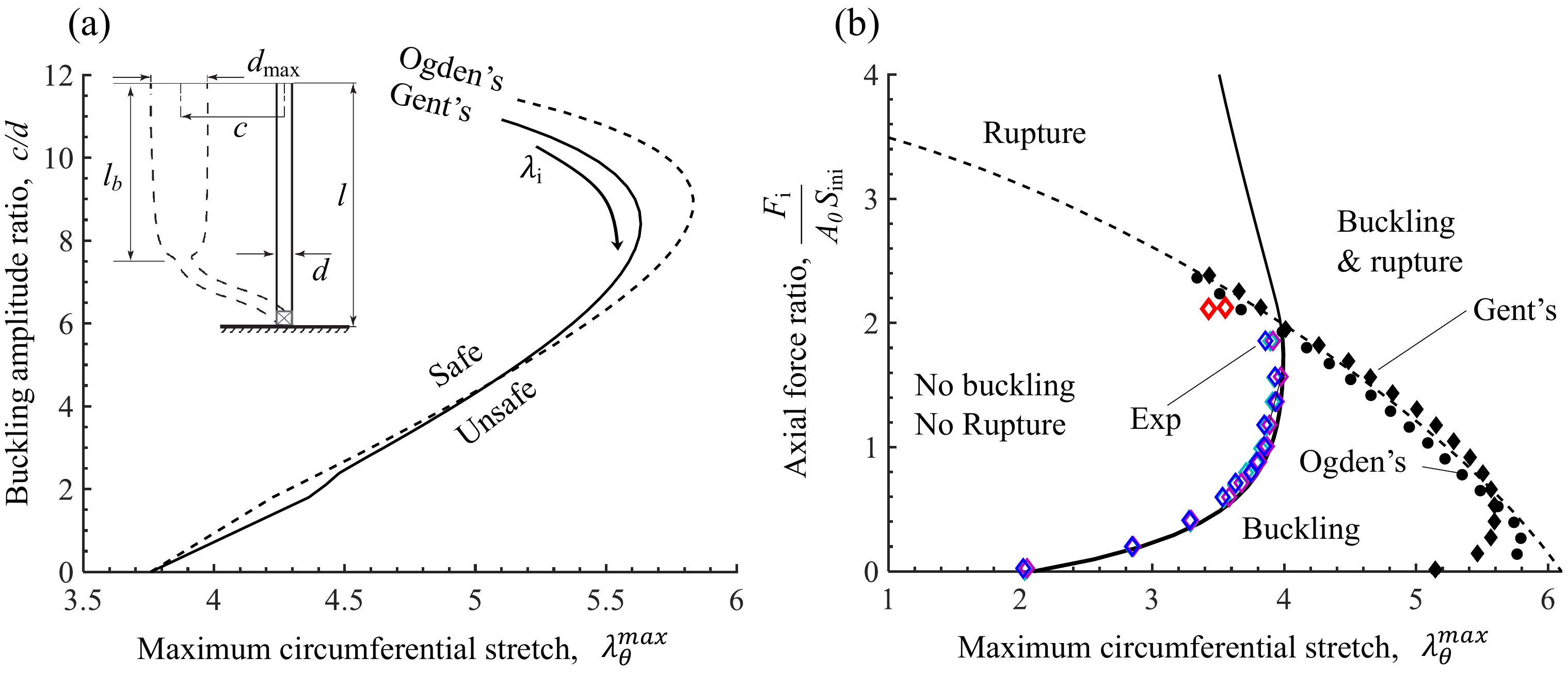}
	\caption{Rupture behaviour of the rubber tube subjected to an internal pressure and fixed longitudinal pre-stretch, $F_{\text{i}}$ (longitudinal tension at the beginning of the inflation), $A_0$ (tube cross section area in deformed configuration). (a) Rupture contour line in the plane of buckling amplitude and maximum circumferential stretch based on Ogden's maximum strain energy density and Gent's maximum stretch invariant. (b) Rupture and buckling phase portrait showing the effect of pre-stretch and buckling on the rupture behaviour.}
	\label{Fig_5}
\end{figure}

A rupture portrait based on buckling amplitude ratio ($c/d$) and maximum circumferential stretch ($\lambda_{\theta}^{\text{max}} = d_{\text{max}}/d$) is shown in~\fref{Fig_5}(a). Two rupture criteria, of Ogden and Gent, are compared. Using FEA, we first obtained the  Gent’s first invariant stretch,~\req{gent}, and Ogden’s SEF,~\req{ogden}, for all the data points throughout the inflation of the tube with different pre-stretch values ($1.75 \leq \lambda_{i} \leq 1.4$). Then, the contour lines, shown in~\fref{Fig_5}(a), have been obtained for all the data points to delineate the safe-unsafe boundary with respect to rupture. Along these curves the pre-stretch varies as indicated in~\fref{Fig_5}(a).  By increasing the pre-stretch, the buckling amplitude falls as well as the maximum circumferential stretch. With a large enough pre-stretch (initial tension, $F_i$), rupture without buckling (see the diamond data points in~\fref{Fig_5}(b)) occurs.  Analytical buckling  and rupture envelopes are obtained by following~\citep{hejazi2021}.  A modified post-buckling rupture line based on FEA is shown for both Gent’s stretch invariant and Ogden’s strain energy density based on the contour lines in~\fref{Fig_5}(a). At higher values of pre-stretch, Ogden's criterion is conservative, and at lower pre-stretch values Gent's criterion is conservative. This is due to the additional tensile strain energy induced by buckling.  %Consequently, this results in a non-monotonic contour line with a larger buckling amplitude and a smaller circumferential stretch at low values of pre-stretch.

% With a low value of pre-stretch (small $F_i$), we see a deviation between FEA and the analytical model, where a large buckling amplitude appears. 

\subsection{GOH material model}

We implemented the GOH material model from~\req{holz} with the  parameters for the control case are given in~\tref{Table_1}.  

\begin{figure}[!h]
	\centering
	\includegraphics[width=\textwidth]{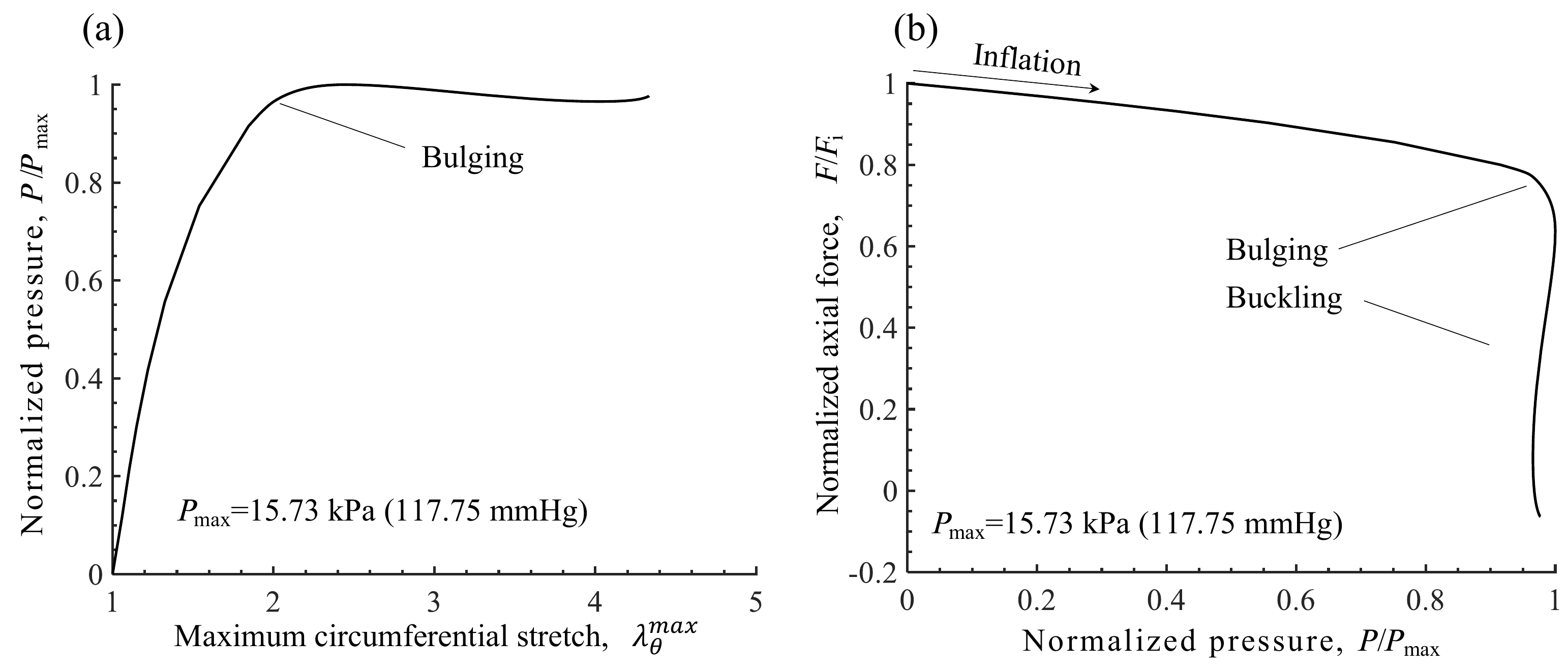}
	\caption{Inflation of a tube with arterial material properties based on fiber reinforced hyperelastic GOH model, showing the onset of bulging and buckling. (a) In contrast to a rubber tube, the internal pressure drop after bulge formation is not significant during the propagation phase. (b) The pressure-force curve shows the onset of bulging (softening behaviour of the pressure ($P$)) and onset of buckling (significant drop in the axial force ($F$)). Here, buckling follows the formation of the aneurysm.}
	\label{Fig_7}
\end{figure}

Comparing the inflation tests on a GOH tube shown in~\fref{Fig_7}(a) with~\fref{Fig_2}(a) for a hyperelastic tube, it can be observed that the decrease in pressure after the formation of the aneurysm is less pronounced for GOH. Hence, the propagation pressure for the aneurysm to spread along the length of the artery is lower. The two instability modes, bulging and buckling can be identified in the pressure-force curve in~\fref{Fig_7}(b). Sudden drops in pressures and axial force follow the bulging and buckling, respectively. 

%be associated wt This is indicated by a  sudden drop of the axial force in~\fref{Fig_7}(b). Here, buckling follows the formation of the aneurysm.  This need not be the case, if we vary the material parameters and/or pre-stretch.

\begin{figure}[!h]
	\centering
	\includegraphics[width=\textwidth]{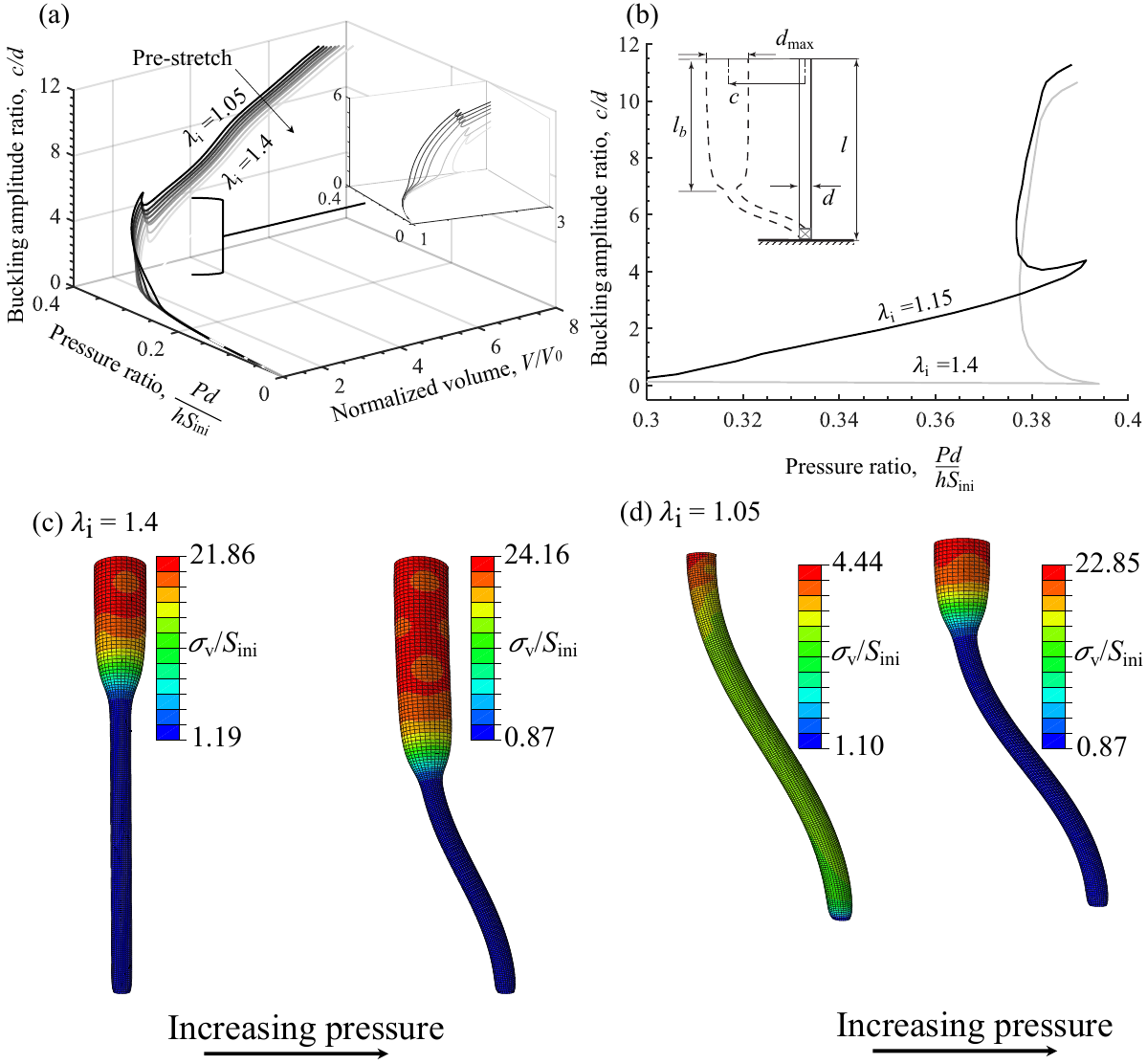}
	\caption{Arterial tube buckling and bulging bifurcation points (limit point instabilities) in the plane of inflation volume versus internal pressure: $P$ (internal pressure), $V$ (internal volume), $V_0$ (internal volume at the beginning of inflation), $S_{\text{ini}}= \displaystyle\frac{1}{2} \left[ \displaystyle\frac{\partial}{\partial \lambda_\theta}\left(\displaystyle\frac{\partial W}{\partial \lambda_z}\right) \right]$ is the initial shear modulus. (a) The bulge and buckling initiation are evident by increase in ($c$/$d$ and limit point instability of $Pd$/$hS_{\text{ini}}$. (b) The effect of tension on the buckling and bulge initiation indicates that with lower pre-stretch tube may buckle before bulge initiation, but as the pre-stretch increases, bulge forms before buckling. Similar to rubber tubes, the propagation pressure does not change due to a change in the pre-stretch. (c) Post bulge buckling of the arterial tube with high pre-stretch. (d) Post buckling bulge of the arterial tube with low pre-stretch.}
	\label{Fig_8}
\end{figure}

A 3D bifurcation map for the tube with the aortic wall is shown in~\fref{Fig_8}(a). %This can be contrasted with~\fref{Fig_3} for a hyperelastic tube. 
In GOH model, compressive forces may arise in the straight regions of the tube without a bulge formation due to stretch inversion.   Buckling before bulging can occur, which is not possible in a hyperelastic tube. In~\fref{Fig_8}(b), we can see a small drop in the pressure as the indicator of the bulge formation and an immediate increase in buckling amplitude indicating the lateral buckling. By comparing two different cases with small ($\lambda_i = 1.15$) and large ($\lambda_i = 1.4$)  pre-stretches, we see how the order of bulging and buckling may reverse (see also FEA results in~\fref{Fig_8}(c) and (d)). 
%With a small pre-stretch, the buckling forms prior to bulge formation and at the onset of bulge formation, the buckling amplitude declines slightly as a precursor to bulging. This is because of the relaxation of the pressure at the onset of bulge localization. During the bulge propagation phase, the buckling amplitude continues its growth. For a large pre-stretch, the axial tension is maintained and buckling is avoided until the bulge forms and propagates leading to the formation of the compression region in the uniform section which produces buckling., similar to a hyperelastic tube. Thus stretch inversion decides the order of buckling or bulging first.

\begin{figure}[!h]
	\centering
	\includegraphics[width=\textwidth]{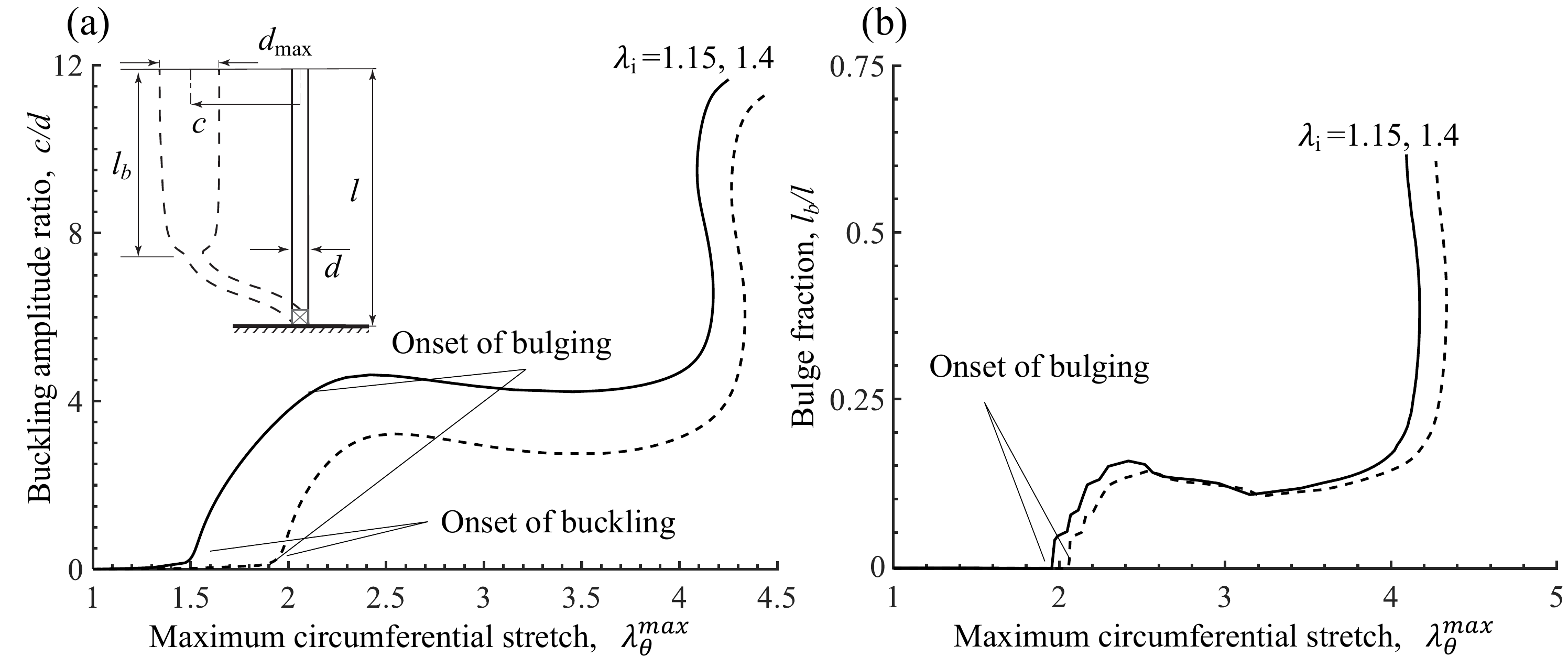}
	\caption{Growth of a bulge in an arterial tube. $\lambda_{\theta}^{\text{max}}=d_{\text{max}}$/$d$ (radial size), $l_b$/$l$ (axial size), $c$/$d$ (lateral buckling growth). (a) Bucking before ($\lambda_{\text{i}}$ =1.15) and after ($\lambda_{\text{i}}$ =1.4) bulge formation. The buckling amplitude ($c$) slightly reduces during propagation phase due to stress relaxation in the uniform section. (b) Initial bulge with a long wave length transforms into a localized bulge and results in a slight reduction in the bulge length ($l_b$). At the late stages of propagation, the axial growth takes over radial growth.}
	\label{Fig_9}
\end{figure}

The growth rate of an AAA is one of the important factors in clinical decision making. Here, we investigated the axial and radial growth and the effect of buckling on these two parameters. \fref{Fig_9}(a) shows the progression of the buckling amplitude ($c$) versus the bulge radial growth ($\lambda_\theta^{\text{max}}$). We see both cases of post bulge buckling ($\lambda_i = 1.15$)  and post-buckling bulge ($\lambda_i = 1.4$). In both cases, the initial stage of the buckling is associated with a surge in amplitude ($c$), followed by a small decrease, and another increase of the amplitude. The buckling and long-wavelength bulge initiate concurrently at the early stages. By the formation of the localized bulge, the buckling amplitude declines slightly, followed by significant radial growth of the bulge. With the transition of radial growth into axial propagation, the buckling amplitude increases further until it reaches the hardening as a result of the reduction in uniform length and pinching effects at the end. Also, with a longer bulge length, the maximum diameter tends to shrink as a result of relaxation of the axial force due to buckling. The axial propagation of the bulge ($l_b$) is shown in~\fref{Fig_9}(b). In the first stage, we see the formation of the long-wavelength bulge. The formation of the localized bulge results in a minor reduction in the bulge length ($l_b$) and a significant increase in radial size. In the next stage of the propagation, the maximum radial size approximately remains constant (marginal reduction is observed) while the bulge propagates axially. 

\begin{figure}[!h]
	\centering
	\includegraphics[width=\textwidth]{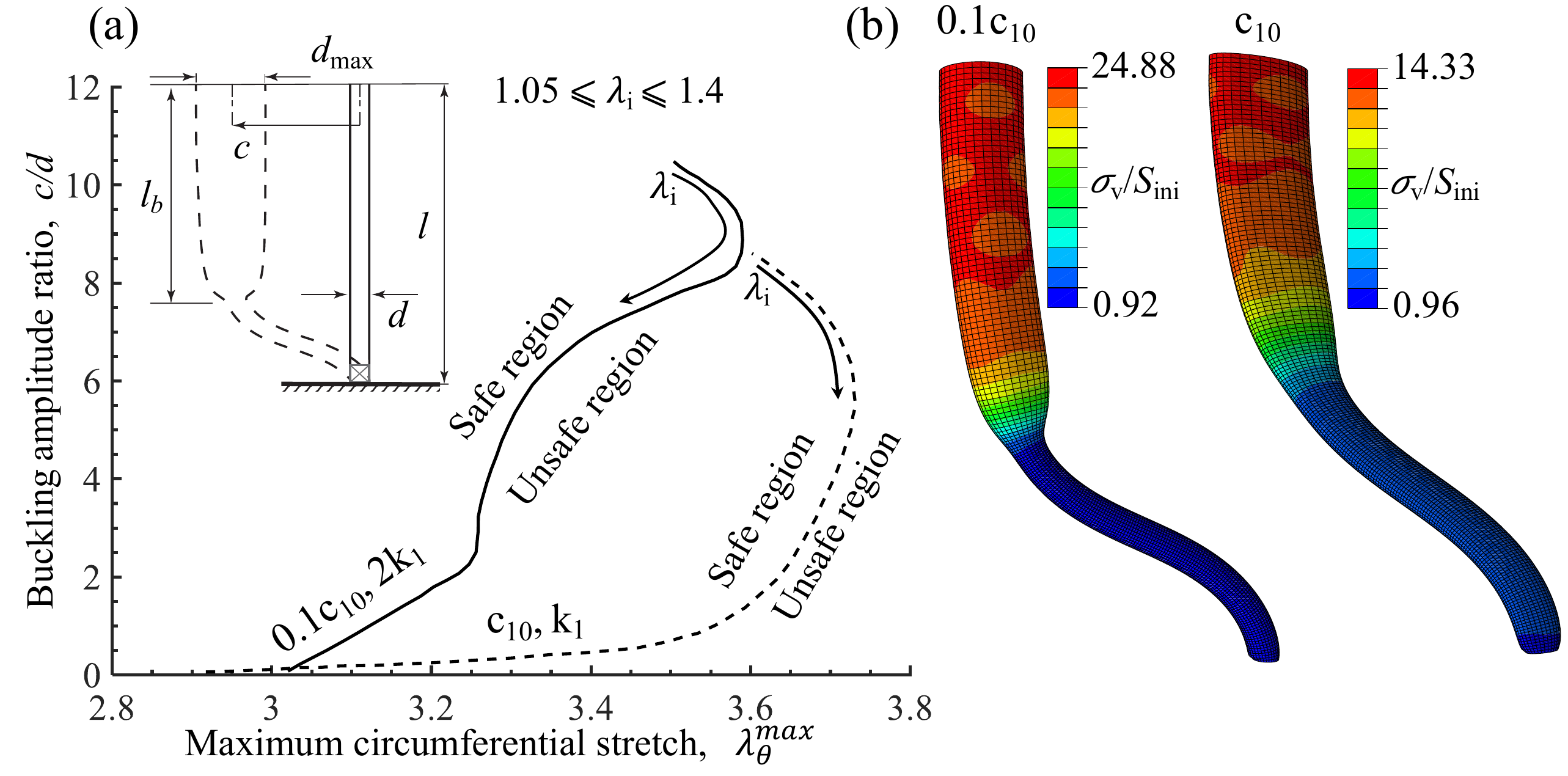}
	\caption{Rupture map for an arterial tube. (a) The effect of material parameters on the rupture of bulged-buckled arterial tube in the plane of maximum radial size and buckling amplitude. (b) Corresponding stress distribution figures for the rupture contours in (a).}
	\label{Fig_10}
\end{figure}

Finally, the rupture map of a tube with the aortic wall with GOH material model, subjected to axial pre-tension and internal pressure is shown in~\fref{Fig_10}(a). By reducing the $c_{10}$ and increasing $k_{1}$, we can model the elastin degradation, which is commonly observed for AAAs~\citep{niestraska2019}.  We showed two rupture thresholds by changing these parameters ($c_{10}$, $k_{1}$) parameter in the SEF function. The rupture criterion is based on the maximum Cauchy stress, which is located at the crown of the bulge on the outer wall with respect to buckling curvature. Similar to~\fref{Fig_5}(a), we obtained the failure points using FEA for various pre-stretch values and sketched the failure boundary on buckling amplitude-circumferential stretch plot. Both of these parameters are easy to measure using common imaging techniques in clinical practice (e.g. CT scans and ultrasound). The control case has a qualitatively similar failure envelope compared to the hyperelastic tube in~\fref{Fig_5}(a). With a larger buckling amplitude for low pre-tension tubes, a smaller circumferential stretch is needed for rupture to occur. However, in general,  buckling acts as a  fail safe mode, and postpones rupture.  When buckling occurs, rupture takes place at a much larger size of the aneurysm, as shown in~\fref{Fig_10}(a). The exponential stiffening is more pronounced at lower values of $c_{10}$ and higher values of $k_{1}$, which prevents the radial growth of the aneurysm and favours buckling. This is also evident in~\fref{Fig_10}(b) which shows higher amplitudes of buckling (lower size of aneurysm) for decreasing $c_{10}$ and increasing $k_{1}$.

%If the pre-stretch is large enough, the tube may rupture without buckling. In the case of tubes with more contribution of fiber reinforcement, the midrange pre-tension values, the rupture is more sensitive to buckling amplitude. By increasing the fiber reinforcement, the stiffness of the wall increases by increasing the radius. Accordingly, the flexural stiffness increases in the bulged region and results in larger stress due to bending.  Also, due to exponential stiffening, the rate of radial growth declines.

%%%%%%%%%%%%%%%%%%%%%%%%%%%%%%%%%%%%%%%%%%%%%%%
%%%%%%%%%%%%%%%%%%%%%%%%%%%%%%%%%%%%%%%%%%%%%%%
\subsection{Clinical implications}
%Buckling is fail safe. Tortuous aneurysms have a larger size without rupturing. One does not know whether buckling or bulging occurs first in real blood vessels due to stretch reversal, which is sensitive. Straight aneurysms are more prone to rupture and hence more frequency screening may be needed. ongoing work with sally.

In this study, we showed two mechanisms that may govern the arterial buckling. The first one, buckling without bulge, is a result of stretch inversion, which is sensitive to microscopic fiber orientation and composition. The second mechanism, probably more relevant to AAAs, is the formation of the compressive force due to axial propagation of the aneurysm. It is not clear whether the buckling always follows the bulge or vice versa. However, many AAAs are commonly observed in a buckled state~\citep{fillinger2004,boyd2021,hejazi2021} do not rupture at a much larger size~\citep{sall2021}. This study suggests that  a straight aneurysm has a higher risk of rupture compared to a buckled aneurysm.
\subsection{Limitations}
There are limitations of this study. The most significant is the lack of haemodynamics and fluid-structure interactions. Such formidable calculations need to be attempted in the future, specially when the artery undergoes a structural instability like buckling. Arterial branches and  the curvature imposed by spine posterior support are not considered. Nonetheless, the rupture maps for both the rubber tube and the tube with an aortic wall can provide useful information regarding the rupture of tortuous AAAs by correlating the 2D buckling amplitude and maximum diameter to the rupture risk. We anticipate such studies in the future.

\section{Conclusion}

We performed a series of inflation tests on rubber tubes with axial pre-tension under fixed-fixed boundary conditions. We performed finite element calculations using ABAQUS$^\copyright$ commercial solvers including linear perturbation analysis and Riks method solver. The FEA code has shown reasonable agreement with \emph{in vitro} experiments on hyperelastic rubber tubes. With a validated FEA code, we implemented the properties of the aortic wall using fiber-reinforced hyperelastic material with collagen and elastin fibers.  We investigated the bulging and buckling instabilities, and explored the bulge growth and rupture in post bulge/buckling regime. We implemented different AAA pathogenesis scenarios by changing the material parameters associated with fiber deposition and degradation. We presented a rupture map in the plane of buckling amplitude plotted against maximum circumferential stretch, both of which can be measured using 2D clinical images (CT and ultrasound) with minimal technical requirements (e.g. 3D image reconstruction). We concluded that the post bulge buckling is a protective mechanism against rupture. 
%%%%%%%%%%%%%%%%%%%%%%%%%%%%%%%%%%%%%%%%%%%%%%%
%%%%%%%%%%%%%%%%%%%%%%%%%%%%%%%%%%%%%%%%%%%%%%%
\section*{Acknowledgment}
We gratefully acknowledge the funding provided for this research by the National Science and Engineering Research Council of Canada (NSERC) through Discovery Grant.  Discussions with Prof.~Y.~Hsiang from the Vancouver General Hospital are thankfully acknowledged.
%\section*{References}
\bibliography{mhsp}
\clearpage
\pagenumbering{gobble}
%\listoffigures
\end{document}